%% file: Multi_dim_der_free_opt_hfr.tex
\documentclass[11pt]{fizik}
\usepackage{hyperref}
\hypersetup{
colorlinks=true,
urlcolor=blue,
citecolor=blue}
\usepackage[all]{xy,xypic}
\usepackage{amsfonts,amssymb,amsmath,amsgen,amsopn,amsbsy,theorem,graphicx,epsfig}
\usepackage{eufrak,amscd,bezier,latexsym,mathrsfs,enumerate,multirow}
\usepackage[utf8]{inputenc}
\usepackage[english]{babel}
\usepackage[dvipsnames]{xcolor}
\usepackage{academicons}
\definecolor{orcidlogocol}{HTML}{A6CE39}
\usepackage[pagewise]{lineno}
\usepackage{epstopdf}
\usepackage{xspace}
\usepackage{color}
\usepackage{units}
\usepackage{slashed}
\usepackage{gensymb}
\usepackage{booktabs}
\usepackage{xcolor}
\usepackage{setspace}
\usepackage{physics}
\usepackage{nicefrac}
\usepackage{bbold}
\usepackage{accents}
\usepackage{mathtools}
\usepackage{dirtytalk}
\usepackage{float}
\usepackage{multirow}
\usepackage{threeparttable}
\usepackage{verbatim}

\def\bt{\begin{equation}}
\def\bea{\begin{eqnarray}}
\def\ee{\end{equation}}
\def\eea{\end{eqnarray}}

\newcommand{\mathdash}{\allowbreak-\allowbreak}

\yil{}
\vol{}
\fpage{}
\lpage{}
\doi{}
\title{
Multidimensional derivative$-$free optimization.\\
A case study on minimization of Hartree$-$Fock$-$Roothaan energy functionals}
\author[BA{\u G}CI]{
\textbf{Ali BA{\u G}CI$^{1}$\thanks{abagci@pau.edu.tr}}
\\
$^{1}$Department of Physics, Faculty of Science, Pamukkale University, Pamukkale, Turkiye\\ [1.8em]

\rec{.. 2025}
\acc{.. 2025}
\finv{.. 2025}
}

\input{fizik.tex}

\begin{document}

\maketitle

\begin{abstract}
This study presents an evaluation of derivative$-$free optimization algorithms for the direct minimization of Hartree$-$Fock$-$Roothaan energy functionals involving nonlinear orbital parameters and quantum numbers with noninteger order. The analysis focuses on atomic calculations employing noninteger Slater$-$type orbitals. Analytic derivatives of the energy functional are not readily available for these orbitals. Four methods are investigated under identical numerical conditions: Powell's conjugate$-$direction method, the Nelder$-$Mead simplex algorithm, coordinate$-$based pattern search, and a model$-$based algorithm utilizing radial basis functions for surrogate$-$model construction. Performance benchmarking is first performed using the Powell singular function, a well$-$established test case exhibiting challenging properties including Hessian singularity at the global minimum. The algorithms are then applied to Hartree$-$Fock$-$Roothaan self$-$consistent$-$field energy functionals, which define a highly non$-$convex optimization landscape due to the nonlinear coupling of orbital parameters. Illustrative examples are provided for closed$-$shell atomic configurations, specifically the $He$, $Be$ isoelectronic series, with calculations performed for energy functionals involving up to eight nonlinear parameters. This work presents the first systematic investigation of derivative$-$free optimization methods for Hartree$-$Fock$-$Roothaan energy minimization with non$-$integer Slater orbitals.

\end{abstract}

\section{Introduction}\label{sec:intro}
In recent years, optimization methods have emerged as essential components within numerous scientific and engineering disciplines, reflecting the increasing interest in their application to minimization of functions, parameter estimation, and model calibration tasks. In many applications, the objective function is nonlinear, high$-$dimensional, or computationally expensive to evaluate. Traditional optimization techniques typically depend on the availability of derivative information, either in the form of exact gradients or numerical approximations. However, for many modern problems, especially those involving simulation$-$based models or black$-$box functions, derivative information may be inaccessible to compute, unreliable, or prohibitively costly. These challenges have motivated the development of derivative$-$free optimization (DFO) methods, which rely solely on objective function evaluations and require no gradient information (see, e.g.,~\cite{1Levis2000, 2Kolda2003, 3Conn2009, 4Rios2013, 5Larson2019} for a detailed overview). A fully rigorous and axiomatic characterization of these methods remains elusive. DFO methods, on the other hand, are particularly well suited for problems involving noisy, non$-$smooth, or discontinuous objective functions, as well as those defined implicitly through complex simulations. The comprehensive survey by Larson et al. \cite{5Larson2019} categorizes DFO techniques into three classes: direct search methods, model$-$based strategies, and stochastic or metaheuristic algorithms. Each is tailored to specific problem structures such as noise levels, dimensionality, and smoothness assumptions.

This study considers the minimization of a real$-$valued function $f:\mathbb{R}^{D}\rightarrow \mathbb{R}$ over a bounded domain $\Omega \subset \mathbb{R}^{D}$, defined by lower $\left(L\right)$ and upper $\left(U\right)$ bounds such that, $L,U \in \mathbb{R}^{D}$ and $L<U$.\\
The set of values is given by,
\begin{align}\label{eq:1}
\Omega=\left\lbrace x \in \mathbb{R} \mid L \le x \le U \right\rbrace.
\end{align}
The objective is to find values $x^{\ast} \in \Omega$ that satisfies the following condition,
\begin{align}\label{eq:2}
f(x^{\ast}) = \min_{x \in \Omega} f(x).
\end{align}
Among the derivative$-$free optimization algorithms \cite{4Rios2013, 5Larson2019, 6Hooke1961}, the algorithm employing Powell’s conjugate direction method \cite{7Powell1962, 8Powell1964, 9Brent1973}, the Nelder$-$Mead simplex algorithm \cite{4Rios2013, 10Nelder1965} and the pattern search algorithm \cite{11Torczon1991, 12Torczon1997}, respectively analyzed, compared. They belong to the class of direct search methods that operate without requiring derivative information, relying entirely on function evaluations. To complement these, a model$-$based derivative$-$free algorithm \cite{13Powell2004, 14Conn2008} is also introduced, which constructs surrogate models of the objective function using radial basis function (RBF) interpolation \cite{15Gutmann2001, 16Jakobsson2010, 17Wild2013}.

The Powell singular function (PSF) \cite{7Powell1962, 18More_1981}, originally introduced in four variables,
\begin{align}\label{eq:3}
f_{4}\left(x\right)=
\left(x_{1}+10x_{2}\right)^{2}+
5\left(x_{3}-x_{4}\right)^{2}+
\left(x_{2}-x_{3}\right)^{4}+
10\left(x_{1}-x_{4}\right)^{4},
\end{align}
with $x=\left(x_{1},x_{2},x_{3},x_{4}\right)^{\intercal} \in \mathbb{R}^{4}$, is a classical benchmark for testing the performance of optimization algorithms. Higher$-$dimensional generalizations are obtained by extending the original four$-$variable structure in a block$-$wise manner, thereby maintaining the fundamental properties of the PSF.\\
For $x \in \mathbb{R}^{D}$, the $D-$dimensional PSF is given by \cite{19Jamil2013},
\begin{align}\label{eq:4}
f_{D}\left(x\right)=\sum_{i=1}^{D/4}
\left(x_{4i-3}+10x_{4i-2}\right)^{2}+
5\left(x_{4i-1}-x_{4i}\right)^{2}+
\left(x_{4i-2}-x_{4i-1}\right)^{4}+
10\left(x_{4i-3}-x_{4i}\right)^{4} .
\end{align}
Despite being continuously differentiable and convex, the Powell singular function presents numerical challenges due to the degeneracy of its second derivatives at the unique minimizer. The Hessian matrix is non$-$singular at typical initial points such as $x^{\ast}=\left(3,-1,0,1,...,3,-1,0,1\right)^{\intercal}$. It becomes however, singular at $x^{\ast}=0$, where several second$-$order partial derivatives vanish. Such features make this function an appropriate benchmark for examining the convergence and robustness of derivative$-$free optimization strategies.

One of the most challenging tasks in quantum mechanical calculations of atoms and molecules is the optimization problem, particularly the optimization of coefficients $C$ \cite{20Roos1980, 21Pulay1980, 22Hamilton1986, 23Francisco2004, 24Yoshikawa2022, 25Sethio2024, 26Lehtola2025} arising in linear combination of atomic orbitals (LCAO) method \cite{27Roothaan1951},
\begin{align}\label{eq:5}
\min_{C} E\left(C\right),
\end{align}
and the orbital parameters $\left(\zeta\right)$ \cite{28Dementev1984, 29Koga1997, 30Koga1997, 31Petersson2003, 32Zijlstra2008, 33Shimizu2011, 34Shaw2023},
\begin{align}\label{eq:6}
\min_{\zeta}E_{SCF}\left(\zeta\right) ,
\end{align}
here, $\zeta=\left(\zeta_{1},\zeta_{2}, ...,\zeta_{D}\right)$ denote the optimization vector for orbital parameters. The coefficients in the LCAO expansion are obtained via self$-$consistent field (SCF) approximation \cite{35Hartree1928, 36Hartree1928}. This approximation iteratively updates the electron density until convergence is achieved. The resulting energy obtained is optimal within the chosen basis set. It does not however, guarantee the global, variational minimum for the electronic energy \cite{21Pulay1980, 26Lehtola2025}. The orbital parameters in the basis functions constitute nonlinear degrees of freedom that must also simultaneously be optimized. In DFO algorithms, the objective function to be optimized corresponds to the total electronic energy depend to nonlinear orbital parameters. Given that, the investigation herein is restricted to atomic systems with few electrons, the standard fixed$-$point SCF procedure, instead of optimization of the linear combination coefficients, is adopted. The aforementioned strategies are analogously applied to the optimization of the orbital parameters in the algebraic solution of the Hartree$-$Fock$-$Roothaan equations \cite{27Roothaan1951, 35Hartree1928, 36Hartree1928, 37Fock1930}, are used to represent the one$-$electron orbitals within the determinant constituting the total electronic wave$-$function.  

Recently, there has been growing interest in approaching the Hartree$-$Fock limit \cite{38Fischer1977} which represents the exact single$-$determinant solution of the many$-$electron electronic Schr{\"o}dinger equation. Achieving the complete basis set limit, though computationally demanding, allows systematic convergence toward this limit. Any sufficiently flexible hydrogen$-$like basis function, with appropriate radial and angular components, will ultimately allow the total electronic wave$-$function to approximate the exact single$-$determinant solution. This has been recently confirmed using Coulomb$-$Sturmians \cite{39Calderini2012, 40Herbst2019, 41Gebremedhin2023}, Lambda \cite{42Hatano2020, 43Hatano2022, 44Hatano2025} and Ba{\u g}c{\i}$-$Hoggan exponential$-$type orbitals (BH–ETOs) \cite{45Bagci2025, 46Bagci2025} basis sets. The wave$-$function can be represented essentially within the relevant Hilbert space, up to numerical precision. Accordingly, the optimization of orbital parameters can now be considered an effective strategy for accelerating convergence toward the Hartree$-$Fock limit. One can approximate the exact solution with a significantly smaller number of basis functions by optimizing the orbital parameters through the Hartree$-$Fock energy functional, thereby reducing computational cost while retaining accuracy.

Most quantum chemistry software packages \cite{26Lehtola2025,34Shaw2023} (see also references therein) employ Gaussian$\mathdash$type orbitals due to their computational efficiency in evaluating molecular integrals. These orbitals allow for analytical calculation of multi$-$center integrals, which significantly reduces computational cost, especially for large molecules. Compared to exponential$-$type orbitals, however, they do not satisfy the electron$-$nuclear cusp condition \cite{47Kato1957} and decay rapidly at long distances \cite{48Agmon1982} which, may reduce accuracy near nuclei or for describing the electron density.\\
Slater$-$type orbitals ($n-$STOs) \cite{49Slater1930}, in contrast, correctly reproduce the cusp at the nucleus and exhibit the proper exponential decay at long range, making them more physically accurate for atomic and molecular wave$-$functions. The main disadvantage of $n-$STOs is that their molecular integrals are computationally demanding and typically require numerical or approximate methods, which limits their practical use in conventional quantum chemistry packages. Consequently, $n-$STOs are predominantly implemented in specialized programs \cite{50Bouferguene1996, 51Rico1998, 52Rico2001} designed for specific applications, yet those implementations often lack general flexibility for orbital optimization within the LCAO framework. The practical implementation of $n-$STOs is thereby challenged by the absence of comprehensive and adaptable optimization protocols. Although $n-$STOs offer clear physical advantages over Gaussian$-$type orbitals, there is a shortage of general and adaptable orbital optimization methodologies compatible with $n-$STOs. Most existing SCF and orbital optimization procedures have been developed with GTOs in mind and cannot be directly applied to $n-$STOs based LCAO calculations. Accurate variational optimization of nonlinear orbital parameters for $n-$STOs continues to pose difficulties. The situation is further complicated for principal quantum numbers with noninteger order, owing to the absence of closed$-$form integral expressions.

The objective of the present work is to investigate multi$-$dimensional derivative$-$free optimization procedures for effective treatment of Slater$-$type orbitals. A systematic investigation of this problem for quantum numbers with noninteger order does not appear to be available in the existing literature. Slater$-$type orbitals are obtained by taking the highest power of $r$ in hydrogen$-$like orbitals. A generalized pre$-$complex form of Slater$-$type orbitals \cite{53Parr1957}, defined by non$-$integer principal quantum numbers \cite{54Zener1930} and known to improve convergence of energy in SCF calculations \cite{29Koga1997, 53Parr1957, 30Koga1997}. From recent theoretical considerations by the author, it follows that Slater$-$type orbitals with non$-$integer quantum numbers originate from the BH$-$ETOs \cite{55Bagci2023, 56Bagci2025} represent a complete orthonormal basis for solution of the quantum mechanical Kepler problem. The hydrogenic bound$-$state solutions of the Schr{\"o}dinger equation constitute a particular subspace in the corresponding Hilbert space.\\
The objective function to be optimized corresponds to the total electronic energy has accordingly, the following form,
\begin{align}\label{eq:7}
E_{min}\left(n^{\ast},\zeta\right)
=\min_{
x
} E_{SCF}
\left(x\right),
\end{align}
here, $x \in \mathbb{R}^{D}$, $x=\left(n^{\ast}_{1}, \zeta_{1},\zeta_{2},...,\zeta_{D-1}\right)$ denote the optimization vector, comprising a principal quantum number of non$-$integer order and a set of orbital parameters. The numerical investigations consider energy functionals with a system$-$dependent number of nonlinear parameters. Computations are performed for closed$-$shell atomic configurations, and results are reported for the $He$, $Be$ isoelectronic series. The number of nonlinear parameters to be optimized is therefore, bounded by eight.

The present paper provides a controlled numerical comparison of DFO methods using identical stopping criteria, precision settings, and atomic test systems, with emphasis on function evaluation counts, CPU time, and convergence characteristics as the dimensionality of the parameter space increases. The organization of the study is as follows. In Section \ref{sec:revisitdfo}, the derivative$-$free optimization strategies are revisited with particular attention to the algorithmic choices and code implementations developed in this work in \textit{Mathematica} programming language. Section \ref{sec:nsto} is devoted to Slater$-$type orbitals with non$-$integer quantum numbers ($n^{\ast}-$STOs). It is shown that $n^{\ast}-$STOs arise naturally when the quantum mechanical Kepler problem is solved. The quantum numbers with non$-$integer order do not originate from fractional calculus or fuzzy logic; rather, the condition for the radial nodes $n^{\ast}-l^{\ast}-\nu \in \mathbb{Z}^{+}$ still holds, where $\left\lbrace n^{\ast},l^{\ast} \right\rbrace \in \mathbb{R}^{+}$ and $0< \nu \leq 1$. The Hartree$-$Fock$-$Roothaan (HFR) equations are presented in Section \ref{sec:dmhfr}, where the explicit form of the total energy functional to be optimized is also given here. The computational results and related discussion are provided in Section \ref{sec:resdis}.
\begin{table}[htp!]
\centering
\caption{Results for four$-$dimensional Powell singular function, comparing different derivative$-$free optimization methods. Here, $N_{f}$ denotes the number of function evaluation, $\left\lbrace x_{i} \right\rbrace$ are the optimized variables.}
\label{tab:1}
\begin{tabular}{ccccc}
\toprule
Method &
$N_f$ &
$Cpu\left(\text{sn}\right)$ &
$\{x_{1},x_{2},x_{3},x_{4}\}$ &
$f_{4}\left(x\right)$
\\
\midrule
\multirow{5}{*}[5.5ex]{Powell CD} &
549 &
1.23 &
\begin{tabular}[c]{@{}r@{}}
$+0.01579\,40598$
\\
$-0.00160\,25513$
\\
$-0.00477\,37965$
\\
$-0.00518\,89743$
\end{tabular} &
\begin{tabular}[c]{@{}r@{}}
$2.85406\,75843 \times 10^{-6}$
\end{tabular}
\\
\midrule
\multirow{5}{*}[5.5ex]{NM Simplex} &
1415 &
1.51 &
\begin{tabular}[c]{@{}r@{}}
$-0.00737\,78023$
\\
$+0.00073\,79311$
\\
$-0.00979\,67719$
\\
$-0.00979\,91776$
\end{tabular} &
\begin{tabular}[c]{@{}r@{}}
$1.26915\,19680 \times 10^{-8}$
\end{tabular}
\\
\midrule
\multirow{5}{*}[5.5ex]{PS$-$C} &
1606 &
0.65 &
\begin{tabular}[c]{@{}r@{}}
$+0.18965\,40391$
\\
$-0.01893\,30248$
\\
$+0.16106\,78404$
\\
$+0.16263\,95424$
\end{tabular} &
\begin{tabular}[c]{@{}r@{}}
$1.06756\,20929 \times 10^{-3}$
\end{tabular}
\\
\midrule
\multirow{5}{*}[5.5ex]{MB$-$RBF} &
453 &
135 &
\begin{tabular}[c]{@{}r@{}}
$+0.01656\,31719$
\\
$-0.00166\,15896$
\\
$+0.02427\,10774$
\\
$+0.02423 64156$
\end{tabular} &
\begin{tabular}[c]{@{}r@{}}
$4.9571460411 \times 10^{-7}$
\end{tabular}
\\

\bottomrule
\end{tabular}
\end{table}
\section{Revisiting the Derivative$-$Free Optimization Algorithms}\label{sec:revisitdfo}
\subsection{Powell’s conjugate direction method}\label{ssec:pcdm}
The Powell’s conjugate direction (Powell’s CD) method \cite{7Powell1962, 8Powell1964, 57Nocedal2006} is useful for optimizing functions that are continuous but not necessarily differentiable. In this method, the search for the minimum is conducted along a set of conjugate directions $\left\lbrace d_{1},d_{2},d_{3},...,d_{D} \right\rbrace$ with respect to a symmetric positive define matrix let say, $Q$. The directions are defined by the condition,
\begin{align}\label{eq:8}
d_{i}^{\intercal}Qd_{j}=0, \forall{i} \neq j.
\end{align}
The Eq. (\ref{eq:8}) meaning that the directions are mutually orthogonal with respect to the matrix $Q$. At iteration $k$ the algorithm initializes the search directions as an $I_{D \times D}$ identity matrix, where $D$ is the dimension of the optimization problem. For $i=1,2,3,...,D$ a line search is performed along $d_{i}$, starting at $x=x_{0}$, where $x_{0}$ is an initial guess with $x_{0} \in \mathbb{R}^{D}$:
\begin{align}\label{eq:9}
\lambda_{i}=\arg\min_{\lambda}
f\left(x_{i-1}+\lambda d_{i}\right),
\end{align}
with $x_{i}=x_{i-1}+\lambda d_{i}$, $\lambda$ is s scalar variable of minimization. The point $x$ is updated by $x_{i} \leftarrow x_{i-1}+\lambda_{i}d_{i}$. Upon completion of the line searches along all $D$ prescribed directions, a new search direction is constructed as $d_{new}=x_{D}-x_{0}$. The algorithm proceeds by updating the direction set through a rotation mechanism; all directions are shifted such that $d_{i} \leftarrow d_{i+1}$ for $i=1,...,D-1$, and the last direction is set to the newly constructed one, $d_{D} \leftarrow d_{new}$. This effectively discards the oldest direction and adds the new conjugate direction. An additional line search is then performed along this new direction to find,
\begin{align}\label{eq:10}
\lambda_{new}=\arg\min_{\lambda}
f\left(x_{D}+\lambda d_{new}\right),
\end{align}
and the initial point for the subsequent cycle is set equal to $x_{0}\leftarrow x_{D}+\lambda_{new} d_{new}$,  after which the iterative process is repeated until convergence. The convergence criterion is satisfied when the absolute improvement in the objective function
\begin{align}\label{eq:11}
\vert \Delta f \vert=
\vert f\left(x_{k}\right)
-f\left(x_{k-1}\right) \vert \leq \epsilon .
\end{align}
Additionally, an early stopping mechanism terminates the optimization if the improvement remains below a threshold for several consecutive iterations, avoiding redundant function evaluations. The implementation incorporates box constraints $x_{i} \in \left[l_{i},u_{i}\right]$ for each variable. Before each line search along direction $d_{i}$ the range of the step parameter $\lambda$ must be determined to ensure that the updated point remains within bounds. After each update, the point is clipped to respect these bounds.\\
The scalar step length $\lambda$ along direction $d_{i}$ is determined using the line search procedure within the computed bounds $\left[\lambda_{min}, \lambda_{max} \right]$. This yields the minimal value of the objective function when restricted to the corresponding one$-$dimensional subspace. Practically, this process is conducted by first identifying an interval containing the minimizer via a bracketing technique, and then refining the solution within this interval using algorithms such as the golden section search or Brent’s method. The golden section search iteratively narrows the bracketing interval let say $\left[a, b\right]$, below tolerance $\epsilon$ based on the golden ratio, ensuring a systematic and reliable convergence towards the minimum within unimodal functions. At each iteration, two interior points are evaluated:
\begin{align}\label{eq:12}
x_{1}&=a+\left(1-r\right)\left(b-a\right)
&
x_{2}=a+r\left(b-a\right),
\end{align}
where, $r=\frac{\sqrt{5}-1}{2} = 1/\varphi \approx 0.618$ and $\varphi=\frac{\sqrt{5}+1}{2}$ is golden ratio. Based on the function values at these points, the interval is reduced if $f\left(x_{2}\right) > f\left(x_{1}\right)$, then $b\leftarrow x_{2}$ and $x_{2} \leftarrow x_{1}$, otherwise, $a\leftarrow x_{1}$, $x_{1} \leftarrow x_{2}$. This process continues until $\vert f\left(x_{1}\right)-f\left(x_{2}\right) \vert < \epsilon$. Brent’s Method \cite{9Brent1973} enhances convergence by integrating parabolic interpolation, in which a parabola is fitted through three points to estimate the location of the minimum, with golden section steps. Both line search methods include error handling and recovery mechanisms to ensure robustness when function evaluations fail.

\subsection{Nelder$-$Mead simplex method}\label{ssec:nmsm}
Instead of moving along a prescribed set of directions as in Powell’s method, the Nelder$-$Mead simplex (NM simplex) method \cite{3Conn2009, 4Rios2013, 10Nelder1965, 57Nocedal2006} performs the search using a simplex, which is a set of $n + 1$ affinely independent points in $\mathbb{R}^{D}$ . For instance, the simplex is a triangle in two dimensions and a tetrahedron in three dimensions.

At iteration $k$ , the simplex is represented as,
\begin{align}\label{eq:13}
\mathcal{S}^{k}=
\left\lbrace
x_{1}^{k}, x_{2}^{k},...,x_{D+1}^{k}
\right\rbrace,
\end{align}
where $x_{i}^{k} \in \mathbb{R}^{D}$. Each vertex corresponds to an evaluation of the objective function $f\left(x_{i}^{k}\right)$. The vertices are ranked so that,
\begin{align}\label{eq:14}
f\left(x_{1}^{k}\right)
\leq
f\left(x_{2}^{k}\right)
\leq
...
\leq
f\left(x_{D+1}^{k}\right),
\end{align}
with $x_{1}^{k}$ being the current best point and $x_{D+1}^{k}$ the worst. The algorithm replaces the worst vertex with a new point that is expected to yield a lower function value. This update is constructed relative to the centroid of the best $D$ vertices (excluding the worst point), defined as,
\begin{align}\label{eq:15}
\overline{x}=\frac{1}{D}\sum_{i=1}^{D} x_{i}^{k} .
\end{align}
The new candidate point is obtained through geometric transformations: reflection, expansion,
contraction, or shrinkage. These transformations are expressed in terms of linear combinations of $\overline{x}^{k}$ and $x_{D+1}^{k}$. The reflection is given by,
\begin{align}\label{eq:16}
x_{r}=\overline{x}^{k}+
\alpha\left(\overline{x}^{k}-x_{D+1}^{k}\right),
\end{align}
where the reflection coefficient is $\alpha = 1.0$. If the reflected point $x_{r}$ is better than the current best, an expansion step is attempted:
\begin{align}\label{eq:17}
x_{e}=\overline{x}^{k}+
\gamma\left(x_{r}-\overline{x}^{k}\right),
\end{align}
with expansion coefficient $\gamma = 2.0$. If reflection does not improve the simplex sufficiently, a contraction is performed:
\begin{align}\label{eq:18}
x_{c}=\overline{x}^{k} \pm
\rho\left(x_{r}-\overline{x}^{k}\right),
\end{align}
with contraction coefficient $\rho = 0.5$. The sign depends on whether an outside or inside contraction is performed. If contraction fails to improve the simplex, a shrinkage operation is applied, moving all vertices toward the best vertex:
\begin{align}\label{eq:19}
x_{i}^{k+1}=x_{1}^{k}+
\sigma\left(x_{i}^{k}-x_{1}^{k}\right),
\quad\quad i=2,...,D+1,
\end{align}
with shrinking coefficient $\sigma=0.25$.\\
To handle bounded optimization domains, all candidate points are reflected into the feasible region using a triangle-wave reflection scheme. This mirrors coordinates that fall outside the bounds back into the valid range. Convergence is assessed using both geometric and function value criteria. The simplex diameter
\begin{align}\label{eq:20}
d^{k}=\max_{i,j} \norm{x_{i}^{k}-x_{j}^{k}}
\end{align}
measures geometric convergence, while the spread in function values
\begin{align}\label{eq:21}
\Delta f^{k}=\max_{i=2,...,D+1}
\left|
f\left(x_{i}^{k}\right)-f\left(x_{1}^{k}\right)
\right|
\end{align}
captures functional convergence. The algorithm terminates when $d^{k}<10^{-8}$ or $\Delta f^{k}<\epsilon$, where $\epsilon$ is the specified accuracy tolerance.\\
To prevent premature convergence, the implementation includes an automatic restart mechanism. When the improvement in the best function value becomes negligible over a specified number of iterations, the algorithm reinitializes the simplex around the current best point with a reduced step size. This anisotropic reinitialization uses edge$-$based scaling derived from the current simplex geometry. Small random perturbations are added to non$-$best vertices to prevent simplex degeneracy.
\subsection{Pattern Search Method: Coordinate/Compass Variant}\label{ssec:prmccv}
The analysis herein is restricted to well-known coordinate-based instances of pattern search, excluding other variants such as Generalized Pattern Search (GPS), MADS, or Hooke$-$Jeeves \cite{2Kolda2003, 5Larson2019, 57Nocedal2006}. In contrast to Powell’s CD method, which relies on successive line minimization along conjugate directions, and the Nelder$-$Mead simplex method, which evolves a simplex through reflection and contraction operations, the coordinate (compass) variant of pattern search advances by systematically polling along coordinate directions with adaptively adjusted step sizes.\\
In the pattern search method \cite{11Torczon1991, 12Torczon1997} with coordinate variant (PS$-$C), the algorithm at iteration $k$ has a current point $x_{k}$ and a step size (or mesh size) $\Delta_{k} > 0$. A finite set of search directions
\begin{align}\label{eq:22}
\mathcal{D}_{k}=
\left\lbrace
d_{k}^{1},...,d_{k}^{m_{k}}
\right\rbrace
\subset \mathbb{R}^{D}
\end{align}
is used to generate trial points,
\begin{align}\label{eq:23}
\mathcal{T}=
\left\lbrace
x_{k}+\Delta_{k}d:d\in \mathcal{D}_{k}
\right\rbrace
\end{align}
The function $f$ is evaluated at these trial points. If any trial point $y \in \mathcal{T}_{k}$ satisfies $f\left(y\right) < f\left(x_{k} \right)$, the algorithm accepts an improvement point and typically expands $\Delta_{k}$. If none improves, $\Delta_{k}$ is reduced and $x_{k+1} = x_{k}$. A set $\mathcal{D} \subset \mathbb{R}^{D}$ is a positive spanning set if the positive cone generated by $\mathcal{D}$ equals $\mathbb{R}^{D}$. Thus, every vector in $\mathbb{R}^{D}$ can be written as a nonnegative linear combination of vectors in $\mathcal{D}$.
 
The implementation supports three pattern types: Coordinate, Compass, and Star. The Coordinate pattern uses
\begin{align}\label{eq:24}
\mathcal{D}=
\left\lbrace
e_{1},e_{2},...e_{D}
\right\rbrace,
\end{align}
the Compass pattern uses
\begin{align}\label{eq:25}
\mathcal{D}=
\left\lbrace
\pm e_{1},\pm e_{2},...,\pm e_{D}
\right\rbrace,
\end{align}
and the Star pattern extends the Compass pattern by additionally including normalized diagonal directions formed from pairs of coordinate directions:
\begin{align}\label{eq:26}
\mathcal{D}=
\left\lbrace
\pm e_{1},\pm e_{2},...,\pm e_{D}
\right\rbrace
\cup
\left\lbrace
\frac{e_{i} \pm e_{j}}{\norm{e_{i} \pm e_{j}}}
: i>j
\right\rbrace,
\end{align}
where $e_{i}$ denotes the canonical unit vectors.\\
The method proceeds according to the following algorithmic steps: polling step, evaluation, acceptance criterion, acceleration, convergence criteria, restart mechanism, stopping rule, respectively.
\begin{itemize}
\item{
Trial points are generated according to Eq. (\ref{eq:23}). The implementation adopts an opportunistic strategy. It first polls along the most recently successful direction (if available) and only explores all remaining directions if this fails.
}
\item{
For each value of $y$, $y \in \mathcal{T}_{k}$, $f\left(y\right)$ is computed. Function evaluations are cached using hash$-$based identifiers derived from rounded coordinate values to avoid redundant computations. In practice, this ensures that function values corresponding to numerically identical trial points are evaluated only once and subsequently recalled.
}
\item{
If there exists a $y \in \mathcal{T}_{k}$ such that $f\left(y\right)<f\left(x_{k}\right)$, the improvement is accepted,
\begin{align}\label{eq:27}
x_{k+1}&=\arg\min_{y \in \mathcal{T}^{k}} f\left(y\right)
&
\Delta_{k+1}=\gamma\Delta_{k}
\end{align}
where the expansion factor $\gamma$ depends on the number of successive improvements, $\gamma = 5.0$ after two or more consecutive successes, and $\gamma = 3.0$ otherwise. If no improvement is found,
\begin{align}\label{eq:28}
x_{k+1}&=x_{k}, & \Delta_{k+1}=\beta\Delta_{k}
\end{align}
where, the contraction factor $\beta$ depends on the number of consecutive failures, $\beta=0.6$ after one failure, $\beta=0.3$ after two failures, and $\beta=0.1$ after three or more failures.
}
\item{
When an improvement is obtained, the algorithm attempts up to three acceleration steps along the displacement direction $d_{accel}=x_{k+1}-x_{k}$ with increasing acceleration factors. The initial factor is $3.0$ increased multiplicatively by $1.8$ after successful acceleration and reduced by a factor of $0.7$ after failure, bounded between $2.5$ and $12.0$.
}
\item{
The algorithm terminates if any of the following conditions is satisfied; $\Delta_{k}<\epsilon_{min}$ ($\epsilon_{min}$ is the minimum step size threshold), the improvement in function value is less then $50\epsilon$ ($\epsilon$ is the accuracy tolerance), the average improvement over the last three iterations is less then $200\epsilon$, four consecutive polling cycles fail to produce improvement, no improvement occurs for ten iterations and the progress rate over a specified interval (default: 25 iterations) falls below a minimum threshold.
}
\item{
If progress fails to improve over a specified interval (evaluated every 25 iterations), the algorithm performs a single restart from the current best point, with the step size reset to $20\%$ of its initial value. At most one restart is permitted.
}
\item{
The algorithm terminates according to the convergence criteria above or when $k \geq k_{max}$. Otherwise, $k \leftarrow k + 1$ and the process returns to the first step.
}
\end{itemize}
\begin{table}[htp!]
\centering
\caption{Results for eight$-$dimensional Powell singular function, comparing different derivative$-$free optimization methods. Here, $N_{f}$ denotes the number of function evaluation, $\left\lbrace x_{i} \right\rbrace$ are the optimized variables.}
\label{tab:2}
\begin{tabular}{ccccc}
\toprule
Method &
$N_f$ &
$Cpu\left(\text{sn}\right)$ &
$\{x_{i}\}_{i=1}^{8}$ &
$f_{8}\left(x\right)$
\\
\midrule
\multirow{5}{*}[5.5ex]{Powell CD} &
6432 &
126.81 &
\begin{tabular}[c]{@{}r@{}}
$+0.00309\,99729$
\\
$-0.00031\,00834$
\\
$+0.00223\,50572$
\\
$+0.00223\,54466$
\\
$+0.00192\,53111$
\\
$-0.00019\,26300$
\\
$+0.00216\,63526$
\\
$+0.00216\,49130$
\end{tabular} &
\begin{tabular}[c]{@{}r@{}}
$9.13902\,55959 \times 10^{-11}$
\end{tabular}
\\
\midrule
\multirow{5}{*}[5.5ex]{NM Simplex} &
1736 &
2.59 &
\begin{tabular}[c]{@{}r@{}}
$+0.00514\,65303$
\\
$-0.00051\,45673$
\\
$+0.00336\,44584$
\\
$+0.00336\,50863$
\\
$-0.00015\,05251$
\\
$+0.00001\,50200$
\\
$-0.00302\,57733$
\\
$-0.00302\,53766$
\end{tabular} &
\begin{tabular}[c]{@{}r@{}}
$1.09927\,97079 \times 10^{-9}$
\end{tabular}
\\
\midrule
\multirow{5}{*}[5.5ex]{PS$-$C} &
25815 &
10.51 &
\begin{tabular}[c]{@{}r@{}}
$+0.10476\,06702$
\\
$-0.01051\,07552$
\\
$+0.07759\,82360$
\\
$+0.07770\,90413$
\\
$+0.11817\,02422$
\\
$-0.01188\,78134$
\\
$+0.08346\,37386$
\\
$+0.08374\,53467$
\end{tabular} &
\begin{tabular}[c]{@{}r@{}}
$1.63408\,67627 \times 10^{-4}$
\end{tabular}
\\
\midrule
\multirow{5}{*}[5.5ex]{MB$-$RBF} &
453 &
728.24 &
\begin{tabular}[c]{@{}r@{}}
$+0.02837\,85339$
\\
$-0.00297\,53226$
\\
$-0.00889\,20821$
\\
$-0.00776\,97728$
\\
$-0.02012\,72767$
\\
$+0.00192\,41878$
\\
$+0.01774\,48486$
\\
$+0.01666\,42711$
\end{tabular} &
\begin{tabular}[c]{@{}r@{}}
$5.02711\,78966 \times 10^{-5}$
\end{tabular}
\\

\bottomrule
\end{tabular}
\end{table}
\subsection{Model$-$Based Method: Radial Basis Function Interpolation}\label{ssec:mbmrbfi}
In derivative$-$free model-based radial basis function (MB$-$RBF) optimization \cite{5Larson2019, 15Gutmann2001, 17Wild2013}, a surrogate model $g\left(x\right)$ is constructed to approximate the true objective function 
$f\left(x\right)$ using a set of previously evaluated sample points,
\begin{align}\label{eq:29}
S=
\left\lbrace
x_{1},x_{2},...,x_{m}
\right\rbrace,
\end{align}
with $m$ is the number of function evaluations (sample points) accumulated so far. This surrogate model is subsequently employed in the generation of new trial points. The conceptual role of $S$ is preserved in the present implementation but the surrogate is constructed from an adaptive subset $\left(S^{\prime}\right)$ consisting of points those with smallest objective values. The radial basis function constructs a surrogate $g\left(x\right)$ that interpolates the true function $f\left(x\right)$ at $\left(S^{\prime}\right)$ to ensure local accuracy controlled by trust$-$region. The dependence of radial basis function interpolation and trust$-$region constraints on Euclidean distances between sample points and candidate points leads to sensitivity with respect to heterogeneous coordinate scaling. Accordingly, surrogate construction and candidate generation are carried out in normalized coordinates on the unit hypercube. They are then, mapped back to the original domain before function evaluation.\\
The surrogate is defined as,
\begin{align}\label{eq:30}
g\left(x\right)=
\sum_{i=0}^{m} \lambda_{i}\phi\left(\norm{x-x_{i}}\right)
+
\sum_{l=0}^{q}\mu_{l}h_{l}\left(x\right),
\end{align}
where, $\phi:\left[0,\infty\right) \rightarrow \mathbb{R}$ is a radial basis function whose values depends on the radial distance. $\left\lbrace \lambda, \mu \right\rbrace$ are coefficients. $h_{l}\left(x\right)$ polynomial basis function of low$-$degree polynomial tail. Its role is to guarantee uniqueness and prevent degeneracy. Exact interpolation is imposed at all sample points,
\begin{align}\label{eq:31}
g\left(x_{j}\right)&=f\left(x_{j}\right) & j=1,2,...,m .
\end{align}
The radial basis function and the polynomial matrices are defined as,
\begin{align}\label{eq:32}
\begin{array}{cc}
\Phi_{ji}=\phi\left( \norm{x_{j}-x_{i}} \right)
\\
H_{jl}=h_{l}\left(x_{j}\right)
\end{array},
\end{align}
where, $\Phi \in \mathbb{R}^{m \times m}$ and $H \in \mathbb{R}^{m \times q}$, with $q$ number of polynomial tail basis terms, respectively. Thus, each interpolation equation becomes,
\begin{align}\label{eq:33}
\left(\Phi \lambda \right)_{j}
+
\left(H \mu \right)_{j}
=
f\left(x_{j}\right).
\end{align}
The uniqueness of the solution is ensured by enforcing the following orthogonality condition $H^{\intercal} \lambda =0$. Finally, the interpolation and orthogonality conditions yield a block linear system as follows,
\begin{align}\label{eq:34}
\begin{bmatrix}
\Phi & H
\\
H^{\intercal} & 0
\end{bmatrix}
\begin{bmatrix}
\lambda
\\
\mu
\end{bmatrix}
=
\begin{bmatrix}
f
\\
0
\end{bmatrix}.
\end{align}
Notice that, no polynomial tail is used. Instead of enforcing uniqueness through polynomial augmentation and orthogonality here, numerical stability is achieved by adaptive diagonal regularization of the radial basis function interpolation matrix and solution by a pseudo$-$inverse. The form of the radial basis function build on $S^{\prime}$ given as,
\begin{align}\label{eq:35}
g\left(x\right)=
\sum_{i=0}^{m} \lambda_{i}\phi\left(\epsilon \norm{x-x_{i}}\right),
\end{align} 
where 
$\epsilon > 0$ is a shape parameter (default $\epsilon=1.5$ in normalized space). Rather than solving Eq. (\ref{eq:34}) directly or forming block system via the polynomial terms, the present approach stabilizes the system via diagonal regularization by,
\begin{align}\label{eq:36}
\left(\Phi+\eta I\right) \approx f,
\end{align}
and computes $\lambda$ using a pseudo$-$inverse. The regularization magnitude $\eta$ is chosen adaptively by defining $\sigma_{max}$ and $\sigma_{min}$ denote the largest and smallest singular values of $\Phi$. The condition estimate is,
\begin{align}\label{eq:37}
\kappa &=\frac{\sigma_{max}}{\sigma_{min}+10^{-20}},
&
\eta =
\begin{cases}
0.1 & \text{if $\kappa > 10^{6}$} \\
0.05 & \text{if $\kappa > 10^{4}$} \\
0.01 & \text{if $\kappa > 10^{3}$} \\
0.005 & \text{if otherwise}
\end{cases}.
\end{align}
The coefficients are then obtained as, $\lambda = \left(\Phi + \eta I \right)^{+} f$.\\
Boundedness of the surrogate is enforced through a clipping operator, preventing the optimization process from exploiting spurious extrema. Empirical lower $\left(L\right)$ and upper $\left(U\right)$ bounds are determined. The clipped surrogate given as,
\begin{align}\label{eq:38}
g_{clip}\left(x\right)=
\min\left\lbrace
U, \max\left\lbrace
L, g\left(x\right)
\right\rbrace
\right\rbrace .
\end{align}
Usually, the next point is generated by a classical trust$-$region formulation, in which a ball of radius centered at the current best point is considered. A candidate point inside this region is selected and evaluated by computing the true objective function value, and the data set is augmented with the pair consisting of the new point and its corresponding function value.  The surrogate model is then rebuilt or updated using the augmented data set. In standard trust$-$region methods, candidate acceptance and the evolution of the radius are based on comparing the reduction predicted by the surrogate model with the reduction obtained in the true objective function. In the present work, this mechanism is replaced by a sampling$-$based approach. Multiple candidate points are generated inside the trust region of radius, the surrogate is evaluated at these points, and the candidate yielding the best surrogate value is selected. Acceptance of the candidate and modification of the trust$-$region radius depend on the improvement achieved in the true objective function.
\begin{figure}[t!]
  \centering
  \includegraphics[width=0.79\textwidth]{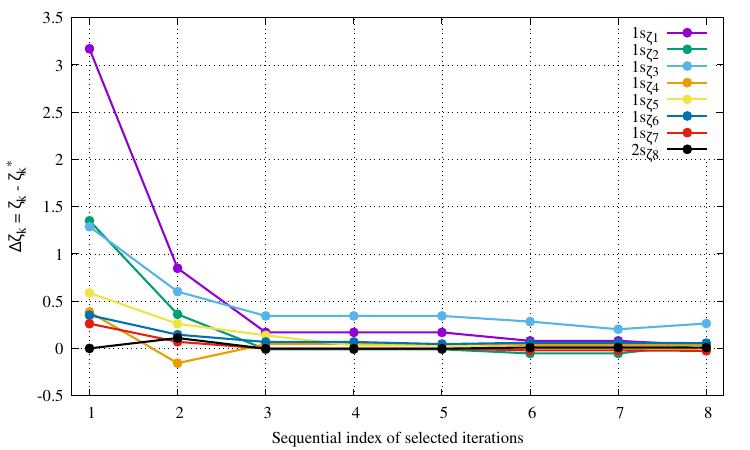}
  \caption{Convergence behaviour of orbital parameters in multidimensional optimization of HFR energy functional for $Be$ atom with NM$-$Simplex algorithm.}
  \label{fig:be_orb_opt}
\end{figure}

\section{Slater$-$type Orbitals with Non-integer Quantum Numbers}\label{sec:nsto}
Infeld and Hull \cite{58Infeld1951} for solution of the Dirac equation \cite{59Dirac1928} and correct representation of its non$-\allowbreak$relativistic limit proposed a generalized Kepler problem. A family of Coulomb$-$like Hamiltonians whose operators preserve the same analytic and symmetry structure as the hydrogenic Hamiltonian were suggested. The generalized Kepler problem extends the standard hydrogen$-$like systems to allow non$-$integer quantum numbers. The Lie algebra structure so(4), so(3, 1) for bound, scattering states, respectively retain their formal character (remain isomorphic under this extension), the introduction of quantum numbers with fractional order represents a controlled deformation of the quantization structure that maintains the integrability properties of the Coulomb potential. The factorization method via recurrence relationships employed in [55] to solve the second order differential equation.

The author in his previous paper \cite{55Bagci2023} suggested an intermediate form between generalized Laguerre and standard Laguerre polynomials. It is referred to as transitional Laguerre polynomials. Instead of using factorization method, this yielded to find exact solution for generalized Kepler problem. The radial part of Bagci$-$Hoggan complete and orthonormal exponential$-$type basis set with non$-$integer quantum numbers, emerge naturally from solution of generalized Kepler problem through the Sturm$-$Liouville formalism were given as \cite{55Bagci2023},
\begin{align}\label{eq:39}
R_{n^{\ast}l^{\ast}}^{\alpha \nu}\left(\zeta, r\right)=
\mathcal{N}_{n^{\ast}l^{\ast}}^{\alpha \nu}\left(\zeta\right)
\left(2\zeta r\right)^{l^{\ast}+\nu-1}
e^{-\zeta r}
L_{n^{\ast}-l^{\ast}-\nu}^{2l^{\ast}+2\nu-\alpha}\left(2\zeta r\right),
\end{align}
with $\left\lbrace n^{\ast},l^{\ast} \right\rbrace \in \mathbb{R}$, $0<\nu \leq 1$, $\zeta$ is the orbital parameter, $\mathcal{N}_{n^{\ast}l^{\ast}}^{\alpha \nu}$ are normalization constants and $\alpha$, $\alpha > -3$ is weighting parameter in Hilbert space, respectively.  BH$-$ETOs form convenient set that spans subspaces of $L^{2}\left( \mathbb{R}^{3} \right)$ appropriate to Coulomb$-$like systems whose effective coupling or geometry may differs from the pure $1/r$. For $\nu=1$, the formalism reduce to the usual hydrogenic solutions. Further details lie beyond the scope of the present paper; interested readers are referred to \cite{55Bagci2023, 56Bagci2025}.\\
$n^{\ast}-$STOs arise from BH$-$ETOs through a simplification by retaining only the highest$-$power term in the Laguerre polynomials.  Four consistent variants of $n^{\ast}-$STOs, corresponding to different sequences of quantum numbers are defined in \cite{55Bagci2023}. The radial node structure, characterized by the number and spatial distribution of spherical surfaces $\left( n_{r}=n^{\ast}-l^{\ast}-\nu \right) \in \mathbb{N}_{0}$, remains invariant for three of them. $n^{\ast}-$STOs, previously postulated in the literature \cite{53Parr1957}, are just one of the variants derived from BH$-$STOs and preserve the radial node condition $n_{r} \in \mathbb{N}_{0}$. They are given by,
\begin{align}\label{eq:40}
\chi_{n^{\ast}lm}\left(\zeta, \vec{r}\right)=
\frac{\left(2\zeta\right)^{n^{\ast}+1/2}}{\Gamma\left(2n^{\ast}+1\right)}
r^{n^{\ast}-1}e^{-\zeta r}S_{lm}\left(\theta, \varphi \right),
\end{align}
$S_{lm}$ are normalized complex $\left(S_{lm} \equiv Y_{lm}, Y^{\ast}_{lm} =Y_{l,-m} \right)$ or real spherical harmonics \cite{60Condon1935}. $\Gamma\left[x\right]$ are the gamma functions.

The one$-$ and two$-$electron atomic integrals used in the HFR calculations over $n^{\ast}-$STOs are given as follows,\\
the one$-$electron integrals,
\begin{align}\label{eq:41}
\bra{p} \hat{O} \ket{q}=
\int \chi^{\ast}_{p}\left( \vec{r} \right)
\hat{O}\left( \vec{r} \right)
\chi_{q}\left( \vec{r} \right)
d^{3}\vec{r},
\end{align}
the two$-$electron integrals,
\begin{align}\label{eq:42}
\bra{pq}\ket{rs}=
\int\int
\chi^{\ast}_{p}\left( \vec{r_{1}} \right)
\chi_{q}\left( \vec{r_{1}} \right)
\hat{O}\left( \vec{r}_{12} \right)
\chi^{\ast}_{r}\left( \vec{r_{2}} \right)
\chi_{s}\left( \vec{r_{2}} \right)
d^{3}\vec{r_{1}}d^{3}\vec{r_{2}}.
\end{align}
Evaluation of atomic integrals given in Eqs. (\ref{eq:41}, \ref{eq:42}) for atomic orbitals with non$-$integer quantum numbers require special attention. These integrals are reduce to hyper$-$geometric functions that are practically difficult to compute. This is due to non$-$analyticity of power series expansion of a function with real exponent \cite{61Weniger2008}. A novel function referred to as Hyper$-$radial functions along with Bi$-$directional method have recently been introduces in \cite{62Bagci2024} by the author. Hyper$-$radial functions allow recurrence relationships for two$-$electron integrals, eliminating the dependence on hyper$-$geometric functions. A detailed treatment for analytical evaluation of atomic integrals over $n^{\ast}-$STOs can be found in \cite{62Bagci2024}.

\section{Direct Minimization of HFR Energy Functional}\label{sec:dmhfr}
The Hartree$-$Fock (HF) method approximates the solution of the many$-$electron Schr{\"o}dinger equation by expressing the electronic wave function as a single Slater determinant composed of spin$-$orbitals. Each spin$-$orbital is a one$-$electron wave function defined as the product of a spatial orbital describing the electron’s distribution in real space and a spin function which specifies its spin state. The resulting wave function is antisymmetric with respect to electron exchange. The atomic Hamiltonian is partitioned into distinct contributions, namely, the one$-$electron terms that include the kinetic energy of each electron, its Coulombic attraction to the nuclei, and the two$-$electron terms that describe electron$-$electron repulsion. The electronic energy is obtained by minimizing the expectation value of the Hamiltonian with respect to the orbitals, leading to a set of SCF equations. Each orbital is determined through an effective one$-$electron operator that accounts for the averaged Coulomb and exchange interactions of all other electrons. Representing spin$-$orbitals as linear combinations of a finite basis set transforms the SCF equations into matrix form, defined by orbital expansion coefficients. In this algebraic representation, the electronic structure problem is defined in terms of these expansion coefficients, yielding a generalized eigenvalue equation that provides a formulation that is suitable for both computational treatment and direct analysis of the associated energy functional.

The spin$-$orbitals are expanded using the LCAO method, where each orbital is written as a weighted sum of localized atomic basis functions. For closed$-$shell atoms, it takes the form:
\begin{align}\label{eq:43}
u_{i}=\sum_{q=1}^{m} \chi_{q}\left( \vec{r} \right) C_{qi},
\end{align}
where, $C_{qi}$ are the orbital coefficients and $m$ is the number of the basis functions. The coefficient matrix $\mathbb{C}= \left[ C_{qi} \right]$ serves as the set of variational parameters that determine the electronic wave function and, consequently, the total energy. The Hartree$-$Fock energy functional can be written as \cite{21Pulay1980, 26Lehtola2025, 27Roothaan1951, 63Szabo1996},
\begin{align}\label{eq:44}
E\left[ \mathbf{C} \right]=
\sum_{i=1}^{N_{occ}}\sum_{pq}2C_{pi}C_{qi}h_{pq}+
\frac{1}{2}\sum_{i,j=1}^{N_{occ}}\sum_{pqrs}
4C_{pi}C_{qi}C_{rj}C_{sj}
\bigg[
\left(pq \vert rs \right)-
\frac{1}{2}
\left(ps \vert rq \right)
\bigg],
\end{align}
$N_{occ}$ is the number of occupied orbitals, $h_{pq}$ are one$-$electron integrals, $\left( pq \vert rs\right)$, $\left( ps \vert rq\right)$ are the two$-$electron Coulomb $\left(J_{pqrs}\right)$, and exchange $\left(K_{pqrs}\right)$ integrals, respectively. The density matrix is defined as,
\begin{align}\label{eq:45}
D_{pq}=2\sum_{i}^{N_{occ}}
C_{pi}C_{qi}.
\end{align}
Using the density matrix the Eq. (\ref{eq:44}) is re$-$written as,
\begin{align}\label{eq:46}
E\left[ \mathbf{C} \right]=
\sum_{pq}D_{pq}h_{pq}+
\frac{1}{2}\sum_{pqrs}D_{pq}D_{rs}
\bigg[
\left(pq \vert rs \right)-
\frac{1}{2}
\left(ps \vert rq \right)
\bigg].
\end{align}
The total electronic energy can accordingly, be written compactly in matrix form as,
\begin{align}\label{eq:47}
E\left[ \mathbf{C} \right]=
Tr\left[ \mathbf{D}\mathbf{h} \right]+
\frac{1}{2}
Tr \bigg[
\mathbf{D} 
\bigg(
\mathbf{J}\left[ \mathbf{D} \right]-
\frac{1}{2}\mathbf{K}\left[ \mathbf{D} \right]
\bigg)
\bigg].
\end{align}
In principle, the values of the energy functional $E\left[\mathbf{C}\right]$ are obtained through the SCF procedure, wherein the linear combination coefficients are $\mathbf{C}$ determined iteratively. $E\left[\mathbf{C}\right]$ may also be evaluated by direct minimization with respect to $\mathbf{C}$, treating the HFR equations as a nonlinear optimization problem. In either case, the solution is constrained such that the resulting orbitals satisfy the orthonormality condition. Note that, one may consider to introduce Lagrange multipliers $\Lambda$ and form the Lagrangian as,
\begin{align}\label{eq:48}
\mathcal{L}\left(\mathbf{C}, \Lambda \right)=
E\left[\mathbf{C}\right]-
Tr \bigg[
\Lambda
\bigg(
\mathbf{C}^{\intercal} \mathbf{S} \mathbf{C}-\mathbf{I}
\bigg)
\bigg].
\end{align}
The Euler$-$Lagrange equations, defined by the condition $\partial_{C}\mathcal{L}=0$, lead to Roothaan equations. In the direct optimization procedures, on the other hand, the energy functional $E\left[\mathbf{C}\right]$ is minimized numerically without reliance on analytic derivatives. A detailed discussion of this topic lies beyond the scope of the present work. Interested readers are directed to \cite{21Pulay1980, 22Hamilton1986, 23Francisco2004, 24Yoshikawa2022, 25Sethio2024, 26Lehtola2025} for a comprehensive treatment.

In the present work the coefficients $\mathbf{C}$ are determined via SCF procedure, the minimum value of total energy is obtained by optimizing the orbital parameters. In compact matrix form, it is written as,
\begin{align}\label{eq:49}
\left\lbrace
n^{\ast}_{i}, \zeta_{i}
\right\rbrace &=
\min_{
\left\lbrace
n^{\prime \ast}_{i}, \zeta^{\prime}_{i}
\right\rbrace
}
E_{SCF} \left[
\mathbf{C}, n^{\prime, \ast}_{i}, \zeta^{\prime}_{i}
\right],
&
\mathbf{C}^{\intercal} \mathbf{S} \mathbf{C}=\mathbf{I}
\end{align}
DFO methods, such as Powell’s conjugate direction, the Nelder$-$Mead simplex, and pattern search methods, can be employed to sample the parameter space for $E_{SCF}\left[\mathbf{C}, n^{\ast}, \zeta\right]$ without explicit gradient or Hessian information. Each of these procedures iteratively proposes new trial values, evaluates the total energy, and updates the search direction or simplex configuration until convergence to the minimum energy is achieved.
\begin{table}[htp!]
\centering
\begin{threeparttable}
\caption{Ground$-$state energies $\left(E\right)$ for He$-$like ions obtained with a minimal basis set approximation based on $n^{\ast}-$STOs, optimized using different derivative$-$free optimization algorithms. The results are given in atomic units $\left(a.u.\right)$}
\label{tab:3}
\begin{tabular}{ccccccc}
\toprule
Method &
Atom &
$N_f$ &
$Cpu\left(\text{sn}\right)$ &
$n^{\ast}$ &
$\zeta$ &
$E$ 
\\
\midrule
\multirow{5}{*}[-14ex]{Powell CD} &
$He$ &
239 &
524.23 &
0.95505\,73500 &
1.61172\,48872 &
\begin{tabular}[c]{@{}r@{}}
$-02.85420\,84970\,26550$
\\
$-02.85420\,84970\,26459$\,\tnote{a}
\\
$-02.85420\,8497\,\tnote{b}\phantom{0\,00000}$
\\
$-02.84765\,6250\,\tnote{b}\phantom{0\,00000}$
\end{tabular}
\\
\cline{7-7}
& $Be^{2+}$ &
373 &
2048.1 &
0.97849\,34043 &
3.60820\,84680 &
\begin{tabular}[c]{@{}r@{}}
$-13.60433\,41353\,32267$
\\
$-13.60433\,4135\,\tnote{b}\phantom{0\,00000}$ 
\\
$-13.59765\,6250\,\tnote{b}\phantom{0\,00000}$
\end{tabular}
\\
\cline{7-7}
& $C^{4+}$  &
239 &
595.4 &
0.98586\,96336 &
5.60713\,94357 &
\begin{tabular}[c]{@{}r@{}}
$-32.35437\,12869\,85264$
\\
$-32.35437\,12869\,83980$\,\tnote{a}
\\
$-32.35437\,1287\,\tnote{b}\phantom{0\,00000}$ 
\\
$-32.34765\,6250\,\tnote{b}\phantom{0\,00000}$
\end{tabular}
\\
\cline{7-7}
& $O^{6+}$ &
373 &
2187.9 &
0.98947\,89476 &
7.60662\,26672 &
\begin{tabular}[c]{@{}r@{}}
$-59.10438\,90714\,93892$
\\
$-59.10438\,9067\,\tnote{b}\phantom{0\,00000}$
\\
$-59.09765\,6250\,\tnote{b}\phantom{0\,00000}$
\end{tabular}
\\
\cline{7-7}
& $Ne^{8+}$ &
386 &
1217.2 &
0.99161\,97334 &
9.60631\,82238 &
\begin{tabular}[c]{@{}r@{}}
$-93.85439\,94999\,65326$
\\
$-93.85439\,9500\,\tnote{b}\phantom{0\,00000}$
\\
$-93.84765\,6250\,\tnote{b}\phantom{0\,00000}$
\end{tabular}
\\
\multicolumn{7}{p{0.95\linewidth}}{
\footnotesize
Calculations performed with precision 50, accuracy 20, and tolerance $10^{-15}$.}
\\
\midrule
\multirow{5}{*}{NM Simplex}
& $He$ &
185 &
266.0 &
0.95505\,73499 &
1.61172\,48828 &
$-02.85420\,84970\,26549$ \\
& $Be^{2+}$ &
151 &
146.8 &
0.97849\,34001 &
3.60820\,84433 &
$-13.60433\,41353\,32267$ \\
& $C^{4+}$ &
174 &
220.8 &
0.98586\,96151 &
5.60713\,93656 &
$-32.35437\,12869\,85260$ \\
& $O^{6+}$ &
188 &
275.3 &
0.98947\,89476 &
7.60662\,26615 &
$-59.10438\,90714\,93892$ \\
& $Ne^{8+}$ &
144 &
134.6 &
0.99161\,97209 &
9.60631\,80975 &
$-93.85439\,94999\,65313$ \\
\multicolumn{7}{p{0.95\linewidth}}{
\footnotesize
Calculations performed with precision 50, accuracy 20, and tolerance $10^{-8}$.}
\\
\midrule
\multirow{5}{*}{PS$-$C}
& $He$ &
351 &
1701.0 &
0.95505\,74382 &
1.61172\,50709 &
$-02.85420\,84970\,26522$ \\
& $Be^{2+}$ &
373 &
2041.2 &
0.97849\,38068 &
3.60820\,99581 &
$-13.60433\,41353\,29824$ \\
& $C^{4+}$ &
374 &
1962.2 &
0.98587\,00134 &
5.60714\,17366 &
$-32.35437\,12869\,80025$ \\
& $O^{6+}$ &
447 &
3294.1 &
0.98947\,88898 &
7.60662\,20917 &
$-59.10438\,90714\,93671$ \\
& $Ne^{8+}$ &
195 &
293.5 &
0.99281\,94262 &
9.61854\,35323 &
$-93.85425\,90088\,85325$ \\
\multicolumn{7}{p{0.95\linewidth}}{
\footnotesize
Calculations performed with precision 50, accuracy 20, and tolerance $10^{-15}$.}
\\
\midrule
\multirow{5}{*}{MB$-$RBF}
& $He$ &
549 &
6264.6 &
0.95505\,78600 &
1.61172\,15781 &
$-02.85420\,84970\,06132$ \\
& $Be^{2+}$ &
587 & 
7951.0 &
0.97884\,58040 &
3.61043\,65835 &
$-13.60433\,14169\,28319$ \\
& $C^{4+}$ &
600 &
8462.8 &
0.98641\,08425 &
5.61358\,22708 &
$-32.35434\,98175\,91416$ \\	
& $O^{6+}$ &
600 &
8236.7 &
0.99005\,42158 &
7.61529\,23745 &
$-59.10435\,05135\,68396$ \\
& $Ne^{8+}$ &
600 &
7918.6 &
0.99126\,34922 &
9.59841\,65766 &
$-93.85436\,72153\,56879$ \\
\multicolumn{7}{p{0.95\linewidth}}{
\footnotesize
Calculations performed with precision 50, accuracy 20, and tolerance $10^{-15}$.}
\\
\bottomrule
\end{tabular}
\begin{tablenotes}
\footnotesize
\item[a,b] Refs.~\cite{45Bagci2025,64Guseinov2008}.
\end{tablenotes}
\end{threeparttable}
\end{table}
\section{Results and Discussions}\label{sec:resdis}
The primary objective of the present paper is to study the numerical performance of multi$-$dimensional DFO algorithms in the direct minimization of atomic HFR energy functionals involving nonlinear orbital parameters and principal quantum numbers with noninteger order. The analysis is accordingly, focused on optimization problems arising from $n^{\ast}-$STOs, where analytic derivatives are unavailable and conventional gradient based techniques are not readily applicable. The Powell's CD, NM$-$Simplex, PS$-$C and NM$-$RBF algorithms are investigated. In particular, attention is focused on the accuracy, computational cost, and convergence behaviour of these methods as the dimensionality of the parameter space increases. All calculations are performed under controlled and identical numerical conditions in order to enable a consistent comparison based on function evaluation counts, central processing unit (CPU) time.

\begin{table}[htp!]
\centering
\begin{threeparttable}
\caption{Ground$-$state energies $\left(E\right)$ in atomic units $\left(a.u.\right)$ for Be$-$like ions obtained with a minimal basis set approximation based on $n^{\ast}-$STOs, optimized using different derivative$-$free optimization algorithms.}
\label{tab:4}
\begin{tabular}{ccccccc}
\toprule
Method &
Atom &
$N_f$ &
$Cpu\left(\text{sn}\right)$ &
$\{n^{\ast}\}$ &
$\{\zeta_{1},\zeta_{2}\}$ &
$E$ \\
\midrule
\multirow{5}{*}[0.5ex]{Powell CD} &
$Be$ &
515 &
5194.1 &
0.98030\,63847 &
\begin{tabular}[c]{@{}r@{}}
3.60870\,56957
\\
0.94739\,72495
\end{tabular} &
\begin{tabular}[c]{@{}r@{}}
$-14.56239\,95174\,17480$
\\
$-14.56492\,26469\,04510$\,\tnote{a}
\\
$-14.56425\,1723\,\tnote{b}\phantom{0\,00000}$ 
\\
$-14.55673\,9859\,\tnote{b}\phantom{0\,00000}$  
\end{tabular}
\\
\cline{7-7}
& $C^{2+}$ &
685 &
12146.1 &
0.98957\,07721 &
\begin{tabular}[c]{@{}r@{}}
5.60072\,51515
\\
1.82174\,49374
\end{tabular} &
\begin{tabular}[c]{@{}r@{}}
$-36.37406\,64868\,97977$ 
\\
$-36.40130\,80035\,97424$\,\tnote{a}
\\
$-36.39977\,5227\,\tnote{b}\phantom{0\,00000}$
\\
$-36.37032\,6680\,\tnote{b}\phantom{0\,00000}$
\end{tabular} \\
\multicolumn{7}{p{0.95\linewidth}}{
\footnotesize
Calculations performed with precision 50, accuracy 20, and tolerance $10^{-15}$.}
\\
\midrule
\multirow{5}{*}[3.0ex]{NM Simplex} &
$Be$ &
180 &
259.0 &
0.98030\,64291 &
\begin{tabular}[c]{@{}r@{}}
3.60870\,56928
\\
0.94739\,71449
\end{tabular} &
\begin{tabular}[c]{@{}r@{}}
$-14.56239\,95174\,17376$
\end{tabular}
\\
& $C^{2+}$ &
171 &
217.1 &
0.98957\,03438 &
\begin{tabular}[c]{@{}r@{}}
5.60072\,25219
\\
1.82174\,34831
\end{tabular} &
\begin{tabular}[c]{@{}r@{}}
$-36.37406\,64868\,89866$
\end{tabular} \\
\multicolumn{7}{p{0.95\linewidth}}{
\footnotesize
Calculations performed with precision 50, accuracy 20, and tolerance $10^{-8}$.}
\\
\midrule
\multirow{5}{*}[3.0ex]{PS$-$C} &
$Be$ &
678 &
12263.7 &
0.98030\,63575 &
\begin{tabular}[c]{@{}r@{}}
3.60870\,54983
\\
0.94739\,70147
\end{tabular} &
\begin{tabular}[c]{@{}r@{}}
$-14.56239\,95174\,17370$
\end{tabular}
\\
& $C^{2+}$ &
469 &
3968.7 &
0.98957\,09291 & 
\begin{tabular}[c]{@{}r@{}}
5.60072\,61839
\\
1.82174\,51147
\end{tabular} &
\begin{tabular}[c]{@{}r@{}}
$-36.37406\,64868\,97093$
\end{tabular} \\
\multicolumn{7}{p{0.95\linewidth}}{
\footnotesize
Calculations performed with precision 50, accuracy 20, and tolerance $10^{-15}$. Step size from $1$ to $0.6$ for $C^{2+}$.}
\\
\bottomrule
\end{tabular}
\begin{tablenotes}
\footnotesize
\item[a,b] Refs.~\cite{46Bagci2025,65Guseinov2009}.
\end{tablenotes}
\end{threeparttable}
\end{table}
\begin{table}[htp!]
\centering
\begin{threeparttable}
\caption{Ground$-$state energies $\left(E\right)$ in atomic units $\left(a.u.\right)$ for Be$-$like ions obtained with a extended basis set approximation $\left(1s_{\zeta_{1}}1s_{\zeta_{2}}2s_{\zeta_{3}}2s_{\zeta_{4}}2s_{\zeta_{5}}\right)$ based on $n-$STOs, optimized using different derivative$-$free optimization algorithms.}
\label{tab:5}
\begin{tabular}{ccccccc}
\toprule
Method & Atom &
$N_f$ &
$Cpu\left(\text{sn}\right)$ &
$\{\zeta_{1},\zeta_{2}\}$ &
$\{\zeta_{3},\zeta_{4},\zeta_{5}\}$ &
$E$
\\
\midrule
\multirow{5}{*}[0.5ex]{Powell CD} &
$Be$ &
1023 &
40868.8 &
\begin{tabular}[c]{@{}r@{}}
3.47780\,22946
\\
6.38674\,00706
\end{tabular} & 
\begin{tabular}[c]{@{}r@{}}
1.36065\,07114
\\
2.80471\,91207
\\
0.86450\,23072
\end{tabular} &
\begin{tabular}[c]{@{}r@{}}
$-14.57302\,02695\,11371$
\\
$-14.57301\,5\,\tnote{a}\phantom{0000\,00000}$
\end{tabular}
\\
& $C^{2+}$ &
$T$ &
$T$ & 
\begin{tabular}[c]{@{}r@{}}
5.45412\,09833
\\
9.71255\,64222
\end{tabular} &
\begin{tabular}[c]{@{}r@{}}
2.97813\,20121
\\
4.39255\,25106
\\
1.91521\,91696
\end{tabular} &
\begin{tabular}[c]{@{}r@{}}
$-36.40849\,16116\,69514$
\\
$-36.40848\,9\,\tnote{a}\phantom{0000\,00000}$
\end{tabular} \\
\multicolumn{7}{p{0.95\linewidth}}{
\footnotesize
Calculations performed with precision 50, accuracy 20, and tolerance $10^{-8}$.}
\\
\midrule
\multirow{5}{*}[0.5ex]{NM Simplex} &
$Be$ &
410 &
2787.6 &
\begin{tabular}[c]{@{}r@{}}
2.78717\,73394
\\
6.23535\,69851
\end{tabular} & 
\begin{tabular}[c]{@{}r@{}}
1.23049\,78180
\\
3.07696\,02046
\\
0.83159\,25245
\end{tabular} &
\begin{tabular}[c]{@{}r@{}}
$-14.57301\,75379\,84983$
\end{tabular}
\\
& $C^{2+}$ &
418 &
3064.6 & 
\begin{tabular}[c]{@{}r@{}}
4.1320078249
\\
9.2242983560
\end{tabular} &
\begin{tabular}[c]{@{}r@{}}
2.2751309848
\\
4.6699369288
\\
1.8703014196
\end{tabular} &
\begin{tabular}[c]{@{}r@{}}
$-36.40849\,16206\,92622$
\end{tabular} \\
\multicolumn{7}{p{0.95\linewidth}}{
\footnotesize
Calculations performed with precision 50, accuracy 20, and tolerance $10^{-15}$.}
\\

\bottomrule
\end{tabular}
\begin{tablenotes}
\footnotesize
\item[a] Ref. \cite{66Clementi1974}
\item[] {$T$ indicates truncated execution.}
\end{tablenotes}
\end{threeparttable}
\end{table}

The computer program code is written in the \textit{Mathematica} programming language \cite{67Mathematica2024} and developed specifically for the present investigation \cite{68Bagci2025}. The code does not rely on any built$-$in optimization commands or numerical solvers provided by the \textit{Mathematica} environment. All DFO algorithms are expressed explicitly at the algorithmic level, including parameter updates, stopping criteria, and convergence checks. It is therefore, not restricted to \textit{Mathematica} and can be translated without substantial modification of algorithm into other programming languages. Calculations are performed on a personal desktop computer with an Intel Core i7$-$5960X (16 threads) and $32$GB RAM running Fedora Linux $42$ 64$-$bit.

The results for the PSF in four$-$ and eight$-$dimensions are reported in Tables \ref{tab:1} and \ref{tab:2}. These functions are continuously differentiable but possesses a degenerate Hessian at the global minimum, which make them suitable for examining the stability of DFO algorithms in the absence of reliable second$-$order information. In the four$-$dimensional case, Powell CD and the NM$-$Simplex algorithms both converge to small residual values of the objective function. The PS$-$C algorithm exhibits slower convergence and reduced accuracy, while the MB$-$RBF algorithm attains intermediate accuracy at a substantially higher computational cost due to surrogate construction. The computational cost of PS$-$C and MB$-$RBF considerably increases when the dimensionality is augmented to eight, both in terms of function evaluations and CPU time.

Tables \ref{tab:3}, \ref{tab:4} summarize the ground$-$state energies, optimization of parameters for He$-$like and Be$-$like ions with minimal basis set approximation of $n^{\ast}-$STOs. The orbital parameters reported in \cite{45Bagci2025, 46Bagci2025} are adopted to determine the parameter bounds listed in these tables. The bounds are specified as,$\left\lbrace \zeta_{i}-\frac{\zeta_{i}}{2},\zeta_{i}+\frac{\zeta_{i}}{2} \right\rbrace$. The initial values are chosen as $\left\lbrace  \zeta_{i}-\frac{\zeta_{i}}{2} \right\rbrace$. This choice allows the behaviour of the optimization procedure to be accurately determined. Substantial differences in efficiency are observed in Table \ref{tab:3}. The NM$-$Simplex algorithm consistently requires fewer function evaluations and shorter CPU times compared to Powell CD and PS$-$C algorithms. The MB$-$RBF algorithm yields acceptable energies but results in a significantly larger computational cost, which limits its practical usefulness for repeated atomic calculations. Given that, the calculations are extended to higher$-$dimensional parameter spaces in the Table \ref{tab:4}, excluding the MB$-$RBF algorithm.\\ Table \ref{tab:4} presents results for the simultaneous optimization of the non-integer principal quantum number $n^{\ast}$ and multiple orbital parameters $\zeta_{i}$. The increased coupling between nonlinear orbital parameters and noninteger quantum number result in a computationally more challenging optimization problem. The PS$-$C algorithm exhibits reliable convergence but demands significantly higher computational effort. Unlike the earlier tables, in which the MB$-$RBF approach dominated the computational cost, the results reported here indicate that the PS$-$C method constitutes the primary source of computational expense. Subsequent investigations are consequently, carried out without further consideration of the PS$-$C algorithm.

The ground$-$state energies reported in Table \ref{tab:5} are obtained using an extended basis set using the $n-$STOs identical to that employed by Clementi in \cite{66Clementi1974}. The results reported in \cite{66Clementi1974} are improved by just refining nonlinear parameters rather than by systematically enlarging the orbital expansion as in \cite{69Koga1995}. This table shows that Powell CD algorithm requires a large CPU time as the number of parameters to be optimized increases, leading termination of the $C^{2+}$ calculations after satisfactory convergence is achieved. The NM$-$Simplex algorithm, by contrast, maintains its stability and yields converged ground$-$state energies but exhibits marginally inferior performance relative to Powell CD algorithm. Note that, tolerances used for Powell CD and NM$-$Simplex algorithms in the Table \ref{tab:4} are exchanged in the Table \ref{tab:5} for a better comparison.

Another calculation is performed for $Be$ atom using seven $1s$ and a single $2s$ orbitals with eight orbital parameters that are variationally need to be optimized. The orbital configuration of the used one$-$electron basis set is,
\begin{align*}
\left\lbrace
1s_{\zeta_{1}},
1s_{\zeta_{2}},
1s_{\zeta_{3}},
1s_{\zeta_{4}},
1s_{\zeta_{5}},
1s_{\zeta_{6}},
1s_{\zeta_{7}},
2s_{\zeta_{8}}
\right\rbrace .
\end{align*}
The optimization is carried out using NM$-$Simplex algorithm under explicit bound constraints $\left(\zeta_{min} \leq \zeta \leq \zeta_{max} \right)$ given as,
\begin{align*}
\left\lbrace
\begin{aligned}
&\{9.512626, 15.854376\} \\
&\{6.754939, 9.456915\} \\
&\{3.864417, 6.440695\} \\
&\{3.086637, 3.858296\} \\
&\{1.762318, 2.937196\} \\
&\{1.054822, 1.758036\} \\
&\{0.524315, 1.048631\} \\
&\{0.410810, 1.232430\}
\end{aligned}
\right\rbrace .
\end{align*}
The optimization is initialized at the lower bound of the parameter domain $\left(\zeta_{init}=\zeta_{min}\right)$. The resulting ground$-$state energy is $E=-14.573023164 a.u.$, in agreement with the re$-$optimized analytical Hartree$-$Fock values reported by Koga using the same basis set in \cite{69Koga1995} (private correspondence). Figure \ref{fig:be_orb_opt} illustrates the evolution of the nonlinear orbital parameters in terms of their deviations $\left(\Delta \zeta \right)$ from the optimized reference values reported in \cite{69Koga1995}. This figure shows that the NM$-$Simplex algorithm yields a stable and consistent evolution of the nonlinear orbital parameters, leading to optimized values that almost reproduce the established analytical Hartree$-$Fock benchmark for the $Be$ atom.

\section{Conclusion}
The results presented in this work demonstrate that, for direct minimization of HFR energy functionals involving noninteger Slater$-$type orbitals, the Nelder$-$Mead simplex algorithm delivers the most consistent and reliable computational performance among the derivative$-$free optimization methods considered. Powell’s conjugate$-$direction method remains effective for problems of lower dimensionality; however, its performance exhibits increasing sensitivity to tolerance settings and stopping criteria as the size of the parameter space grows.

The derivative$-$free optimization algorithms developed here are readily transferable to programming environments other than Mathematica. The relatively long CPU times mainly reflect the computational overhead inherent to Mathematica, including interpreted execution and high$-$precision symbolic$-$numeric handling. The algorithms presented in this study are thus, well suited for application to larger atomic systems and, potentially, to molecular calculations involving exponential$-$type basis functions (see Eq. (\ref{eq:39})). Moreover, using identical orbital configurations, the present methodology yields improved total energies compared with the classical results of Clementi and Roetti \cite{66Clementi1974}. To the best of our knowledge, a systematic analysis of variational optimization strategies for noninteger Slater$-$type orbitals had not previously been performed; the present study helps to close this gap and provides a benchmark for future developments in this area.

\section*{Acknowledgment}
Acknowledgement and/or disclaimer...

\end{document}

%% file: fizik.tex
\newcommand{\bc}{\begin{center}}
\newcommand{\ec}{\end{center}}

\renewcommand{\phi}{\varphi}